\documentclass[aps,prb,twocolumn,unsortedaddress,showpacs,12p]{revtex4-1}

\usepackage{graphicx}
\usepackage{cancel}

\begin{document}

\preprint{PREPRINT}

\title{Vapor-liquid surface tension of strong short-range Yukawa fluid}

\author{G. Odriozola} \affiliation{Programa de Ingenier\'{\i}a Molecular, Instituto Mexicano del Petr\'{o}leo, Eje Central L\'{a}zaro C\'{a}rdenas 152, 07730 M\'{e}xico D.F., M\'exico.}

\author{M. B\'{a}rcenas} \affiliation{Divisi\'on de Ingenier\'{\i}a Qu\'{\i}mica y Bioqu\'{\i}mica, Tecnologico de Estudios Superiores de Ecatepec, Av. Tenologico S/N, 55210 Edo. de M\'{e}x., M\'exico.}

\author{P. Orea}
\email{porea@imp.mx (corresponding author)}
\affiliation{Programa de Ingenier\'{\i}a Molecular, Instituto Mexicano del Petr\'{o}leo, Eje Central L\'{a}zaro C\'{a}rdenas 152, 07730 M\'{e}xico D.F., M\'exico.}

\date{\today}

\begin{abstract}
The thermodynamic properties of strong short-range attractive Yukawa fluids, $\kappa=10, 9, 8$ and $7$, are determined by combining the slab technique with the standard and the replica exchange Monte Carlo (REMC) methods. A good agreement was found among the coexistence curves of these systems calculated by REMC and those previously reported in the literature. However, REMC allows exploring the coexistence at lower temperatures, where dynamics turns glassy. To obtain the surface tension we employed, for both methods, a procedure that yields the pressure tensor components for discontinuous potentials. The surface tension results obtained by the standard MC and REMC techniques are in good agreement. 
\end{abstract}


\maketitle

\section{Introduction}

The short-range hard-core attractive Yukawa (HCAY) fluid has been widely used as a simple model for testing a variety of theories\cite{Pini99, Paricaud06, Nezbeda07, Weiss07, Haro07, Shiqi04} and as a reference system for modeling the behavior of real fluids, such as colloidal suspensions\cite{Pini02, Wu03} and protein solutions\cite{Caccamo00, Sciortino02,Lekker00}. Its high popularity is due to the fact that it captures the most important features of these systems, such as coexistence and interfacial properties which rule many industrial processes and biological phenomena. The HCAY pair potential is given by

\begin{equation}
\label{potencial}
u(r)=\left\{\begin{array}{ll}\infty, & \mbox{ for $r<\sigma,$}\\
             -\epsilon \frac{\exp[-\kappa (r-\sigma)/\sigma]}{r/\sigma}, & \mbox{ for $ \sigma\leq
             r, $}
             \end{array} \right.
\end{equation}

\noindent where $\kappa$ is the interaction range parameter, $\epsilon$ is its depth, and $\sigma =1$ is the hard core diameter.

The vapor-liquid phase diagrams of the HCAY fluids have been accessed by using different simulation techniques as well as different theoretical approaches. In recent previous works the coexistence properties of the HCAY fluid were reported for a wide interaction range~\cite{Minerva01,Orea07,Orea08,Lemus08,Pini10}. For long range interactions, i.~e. for small $\kappa$, the reported values obtained by simulation and different theoretical approaches of the coexistence properties agree. However, for short-range attractions, there appear strong differences between theory and simulations. Moreover, not even different simulation approaches lead to a good agreement. On the other hand, short-range attractive potentials are specially relevant for modeling protein solutions, i.~e., systems having a key role in biological applications~\cite{Frenkel97,Odriozola11}. Hence, there is an increasing attention on this type of interactions~\cite{Weiss08,Weiss09,Nezbeda11,Chapela10,Jakse10,Benavides10}. For these type of interactions, the vapor-liquid coexistence curve turns metastable since it locates within the vapor-solid coexistence curve~\cite{Chen08,Lekker00,Frenkel07,Prausnitz04,Kumar05}. The simulation of this type of systems is rather difficult for standard simulation techniques. 

Determining the surface tension is by far more computationally demanding than obtaining the coexistence curves. This turns even worse when the potential is extremely short-range and discontinuous. This explains why the vapor-liquid interfacial properties of discontinuous potentials have been rarely studied so far. For that reason there is growing interest in developing new simulation techniques for the efficient computation of the surface tension. In this direction, some researchers implemented clever simulation approaches~\cite{Alejandre99, Errington03, Singh03, Jackson05, Errington07, Bryk07, Miguel08, Jackson11}. These approaches, however, provide a considerable variation over their surface tension results \cite{Miguel08}. Furthermore, there are very little data when the potential becomes strong short-range. A recent work of Singh~\cite{Singh09} deals with this issue. He reported the interfacial properties of strong short-range HCAY fluid with $\kappa=8,9,10$, using Grand Canonical transition-matrix Monte Carlo (GC-TMMC). Anyway, the surface tension data are scarce and restricted to the temperature region close to the critical point. This lack of data at lower temperatures motivated us to implement REMC, which should help sampling and thus providing new data in this region. 

In view of the previous paragraphs, we understand that there is a demand for the improvement of the simulation approaches, specially for short-range interaction potentials and at low temperatures. Hence, we combine the slab technique with the REMC method to study the HCAY fluid for $\kappa=10, 9, 8$ and $7$, and for a wide temperature range. The main purposes of this work are three: First, to show that the methodology that has been proposed to calculate the surface tension for discontinuous potentials\cite{Alejandre99,Orea03} is valid even for strong short-range systems. Second, to show that the vapor-liquid coexistence and interfacial properties can be obtained at lower temperatures by using the REMC technique. Third, to report new thermodynamic properties of the strong short-range HCAY fluid for a broader range of thermodynamic conditions, and to compare the results with those previously reported in the literature when available.

\section{Methods\label{methods}}
In the following subsection we summarizes the interface tension evaluation by the virial route for discontinuous potentials. In the next one, some details are given on the implementation of the REMC technique. 

\subsection{Interface tension calculation}

The methodology employed to calculate the interface tension of discontinuous potentials is given in detail in previous papers~\cite{Orea03, Rendon06}. Its main idea is to obtain the pressure tensor through the derivatives of the potential given by~\ref{potencial}. The derivative of the discontinuous contribution in Eq.~\ref{potencial} is given by 
\begin{equation}
\label{deltar} \frac{du(r)}{dr}=-kT \delta(r-\sigma),
\end{equation}
where $\delta(r-\sigma)$ is a $\delta$-function. The evaluation of this expression can be performed during a MC simulation through
\begin{equation}
\label{deltax} \delta(x) = \frac{\Theta(x) - \Theta(x-\Delta
\sigma)}{\Delta\sigma}, ~~~~~~\mbox{as ~~~~~~$\Delta \sigma
\rightarrow 0$},
\end{equation}
where $\Theta(x)$ is the unit step function: $\Theta(x)=0$ for $x < 0$ and $\Theta(x)=1$ for $x > 0$. In this work the parameter $\Delta$ is given the following values: $\Delta=0.005, 0.010, 0.015, 0.020$ and $0.025$.

For the virial route, the surface tension is calculated by
\begin{equation}
\label{st}
\gamma =\frac{L_z}2\Bigg\{\big<P_{zz}\big> -
\frac 12\big[\big<P_{xx}\big> + \big<P_{yy}\big>\big]\Bigg\},
\end{equation}
where $L_z$ is the length of the simulation box in perpendicular direction to the interfaces. The factor $1/2$ is due to the existence of two interfaces in the system.

\subsection{Replica exchange method and simulation details}

As mentioned, we combine the replica exchange Monte Carlo method~\cite{Geyer91,Lyubartsev92,Marinari92,Hukushima96,Frenkel} with the slab technique~\cite{Chapela77}. The general idea is to simulate $M$ replicas of the original system of interest, being each replica at a different temperature, so that the exchange of microstates among the ensemble cells is allowed (swap moves). In this way, replicas at high temperatures travel long distances in phase space, whereas low temperature systems perform precise sampling at local regions of phase space. Hence, by introducing these swap trials, a particular replica travels through many temperatures allowing it to overcome free energy barriers. Particular ensembles are not disturbed but enriched by the different contributions of the $M$ replicas. This technique allows to easily explore the coexistence regions at low temperature, where other methods freeze. The formal justification of the swap trials implies the definition of an extended ensemble as follows
\begin{equation}
Q_{\rm extended}=\prod_{i=1}^{M} Q_{N V T_i},
\end{equation} 
where $Q_{NVT_i}$ is the partition function of the Canonical ensemble of the system at temperature $T_i$, volume $V$, and particle number $N$. $M$ is the considered number of replicas of the system, which equals the number of different temperatures defining the extended ensemble. $Q_{\rm extended}$ may be sampled by $M$ independent standard $NVT$-MC simulations. However, swap trials can be now introduced between the replicas which improves the sampling of the low temperature ensembles. The acceptation probability for this swap trials (performed between adjacent replicas) is given by 
\begin{equation}\label{accP}
P_{acc}\!=\!min(1,\exp[(\beta_j- \beta_i)(U_i-U_j)])
\end{equation}
where $U_i-U_j$ is the potential energy difference between replicas $i$ and $j$, and $\beta_i-\beta_j$ is the difference between the reciprocal temperatures $i$ and $j$. Adjacent temperatures should be close enough to provide large exchange acceptance rates between neighboring ensembles. In order to take good advantage of the method, the ensemble at the larger temperature must assure large jumps in configuration space, so that the small temperature ensembles can sample from disjoint configurations.

We employed parallelepiped boxes with sides $L_x=L_y=8.0\sigma$ and $L_z=40.0\sigma$ for simulating the vapor-liquid interface of the HCAY fluid. Each of them is initially set with all particles ($N=1000$) randomly placed within the slab which is initially surrounded by vacuum\cite{Chapela77}. The center of mass is placed at the box center. Periodic boundary conditions are set in the three directions. Verlet lists are implemented to improve performance. Simulations were carried out in the vapor-liquid metastable region, so the highest temperature is set below the critical temperature. Other temperatures are fixed by following a geometrically decreasing trend. The replicas are equilibrated by $10^{7}$ MC steps, while maximum displacements are varied to yield acceptance rates close to 30 \%. Long displacements trials are also considered. These displacements are important since they allow large jumps in the vapor phase with relatively large acceptance rates while naturally performing particle transference trials between both phases. The thermodynamic properties are calculated by considering additional $4\times10^{7}$ configurations.

The new data reported with standard MC simulations were made using the same parameters as in previous works~\cite{Orea07,Orea08,Lemus08,Pini10}, i.~e., they were performed in a parallelepiped cell with sides $L_z=40\sigma>L_x=L_y=10\sigma$ and $1500$ particles. It is well known that with these parameters finite size effects are avoided~\cite{Orea05,Malfreyt09,Janecek09}. The new results were obtained with $6\times 10^7$ MC steps to get more accurate values than in previous work~\cite{Orea07}. This new results for $\kappa=7$ are slightly different from those previously reported by Duda {\it et.al.}~\cite{Orea07}. Additional standard MC simulations were also obtained with $N=1000$ and $L_x=L_y=8\sigma$ for several conditions, leading to the same results. This suggests that size effects are less important for strong short-range potentials. 

The critical density and temperature were calculated by using the rectilinear diameters law and the universal value of $\beta=0.325$\cite{Allen}. Our results are given in dimensionless units, as follows: $r^*=r/\sigma$ for distance, $T^*= k_BT/\epsilon$ for temperature, $\rho^*=\rho \sigma^3$ for density, and $\gamma^*= \gamma\sigma^2/\epsilon$ for surface tension.

\section{Results and discussion}

\begin{figure}
\resizebox{0.4\textwidth}{!}{\includegraphics{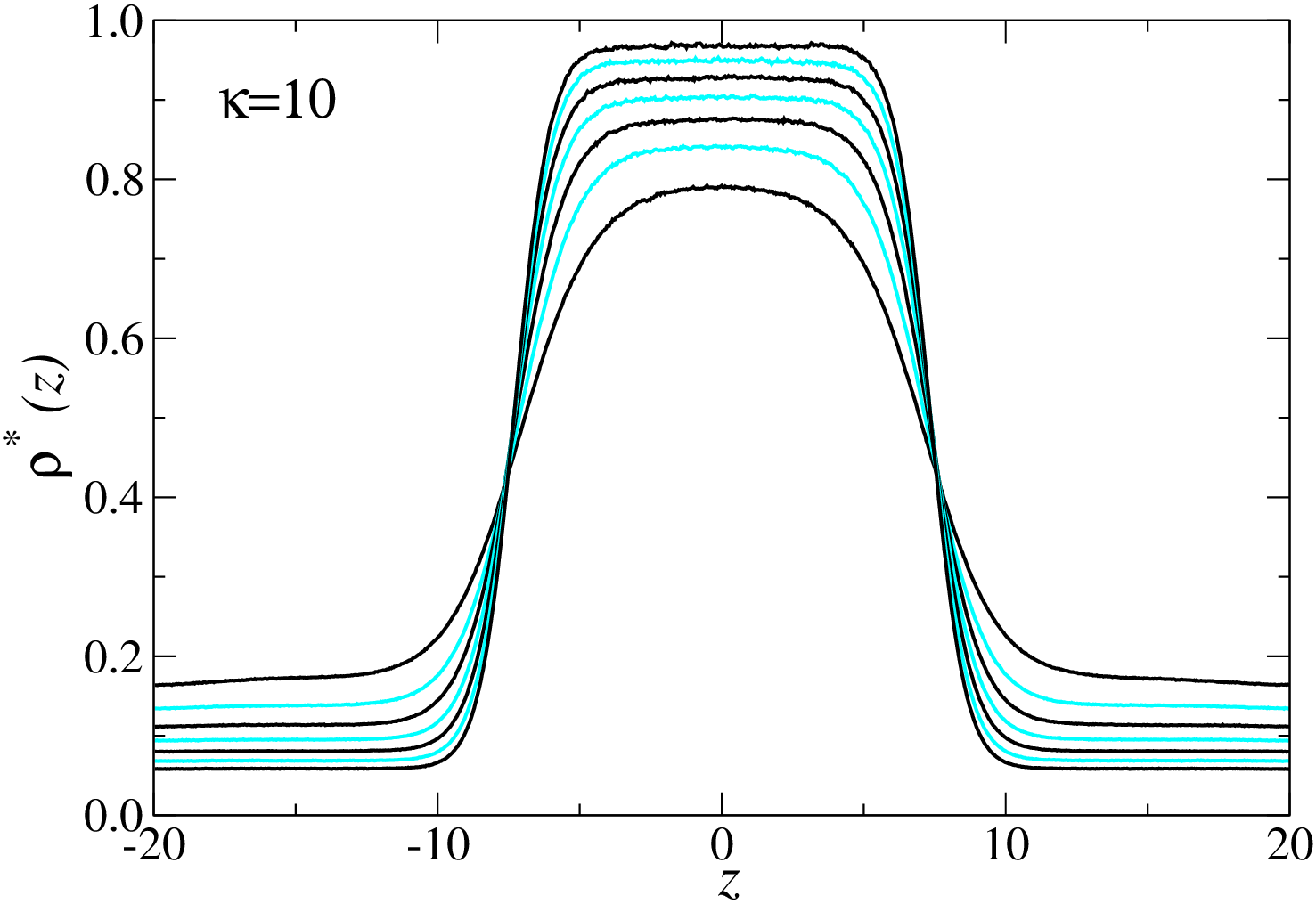}}
\caption{\label{fig1} Density profiles, $\rho^*(z)$, of HCAY fluids with $\kappa=10$ for different temperatures by means of REMC. Larger temperatures produce larger vapor densities and lower liquid densities. }
\end{figure}

The vapor-liquid interfacial properties of strong short-range attractive Yukawa systems with $\kappa=10, 9, 8$, and $7$ were obtained using two simulation techniques, i.~e., standard MC and REMC. Following we focus on the obtained results.

Fig.~\ref{fig1} shows typical interfacial density profiles for strong short-range HCAY fluids with $\kappa = 10$ obtained at different temperatures. Well-defined liquid and vapor regions are observed which makes us confident that the interfaces are stable. Furthermore, the systems show bulk vapor and liquid regions of similar volume. This is convenient to generate relatively large bulk regions and make sure they are fully developed. As it is expected, a temperature increase leads to lower and higher densities of liquid and vapor phases, respectively. Besides, one can clearly see that the interface width increases with temperature, pointing out a decrease of the interface tension. It should be noted that the REMC method yields very smooth density profiles (a good sampling of highly uncorrelated configurations is performed), which allows to obtain precise values of the liquid and vapor coexistence densities, $\rho^*_l$ and $\rho^*_v$, by taking average in the different regions of the profiles. This is done without the need of fitting a hyperbolic function, as commonly used. 

\begin{figure}
\resizebox{0.4\textwidth}{!}{\includegraphics{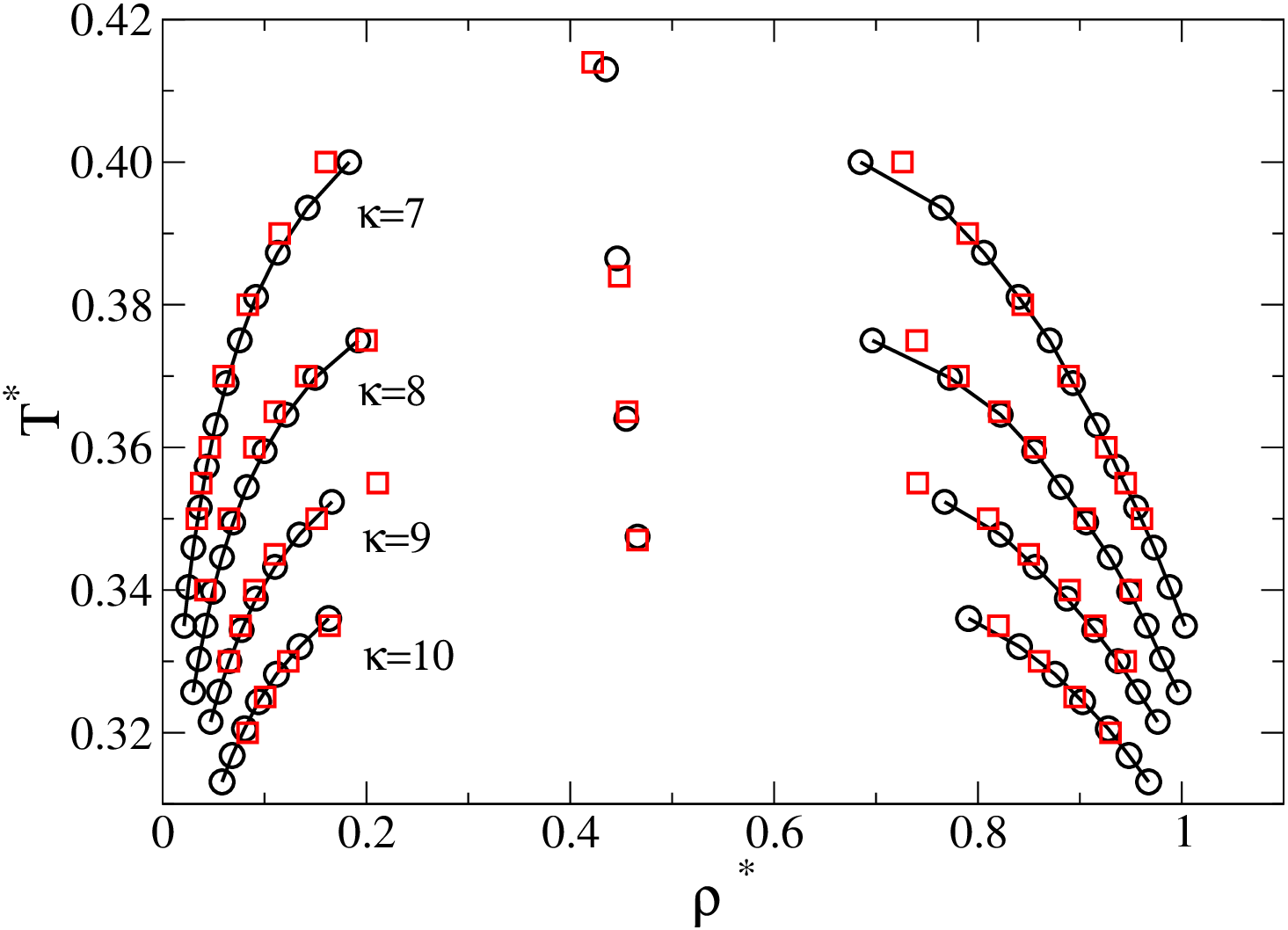}}
\caption{\label{fig2} Phase diagrams of HCAY fluids. Open circles correspond to REMC whereas squares to standard MC. A good agreement is found for all cases.}
\end{figure}

Fig.~\ref{fig2} shows the phase diagrams obtained from REMC for the studied systems (open circles). This figure is built from the values of $\rho^*_l$ and $\rho^*_v$ obtained from the profiles at different temperatures and for $\kappa=10, 9, 8$ and $7$. The same figure also includes previously reported data for $\kappa=10, 9$ and $7$ (open squares)~\cite{Pini10}. The open squares corresponding to $\kappa=8$ are new data. All the data represented with open squares were obtained by using the standard MC method~\cite{Pini10}. An excellent agreement is found between REMC and standard MC (codes are totally independent from one another). The data also agree well with the coexistence curves reported by Singh~\cite{Singh09} (as shown in ref.~\cite{Pini10}). Note that the REMC technique provides data at lower temperature, where the standard methodologies have sampling issues as the system dynamics turns glassy~\cite{Testard11}. At even lower temperatures the vapor-liquid coexistence metastability brakes and crystallization takes place~\cite{Odriozola11} (not shown). For obtaining the critical points we employed the exponential fitting (with $\beta=0.325$)~\cite{Frenkel}. Critical points also show a good match between standard MC and REMC simulations. 

\begin{figure}
\resizebox{0.4\textwidth}{!}{\includegraphics{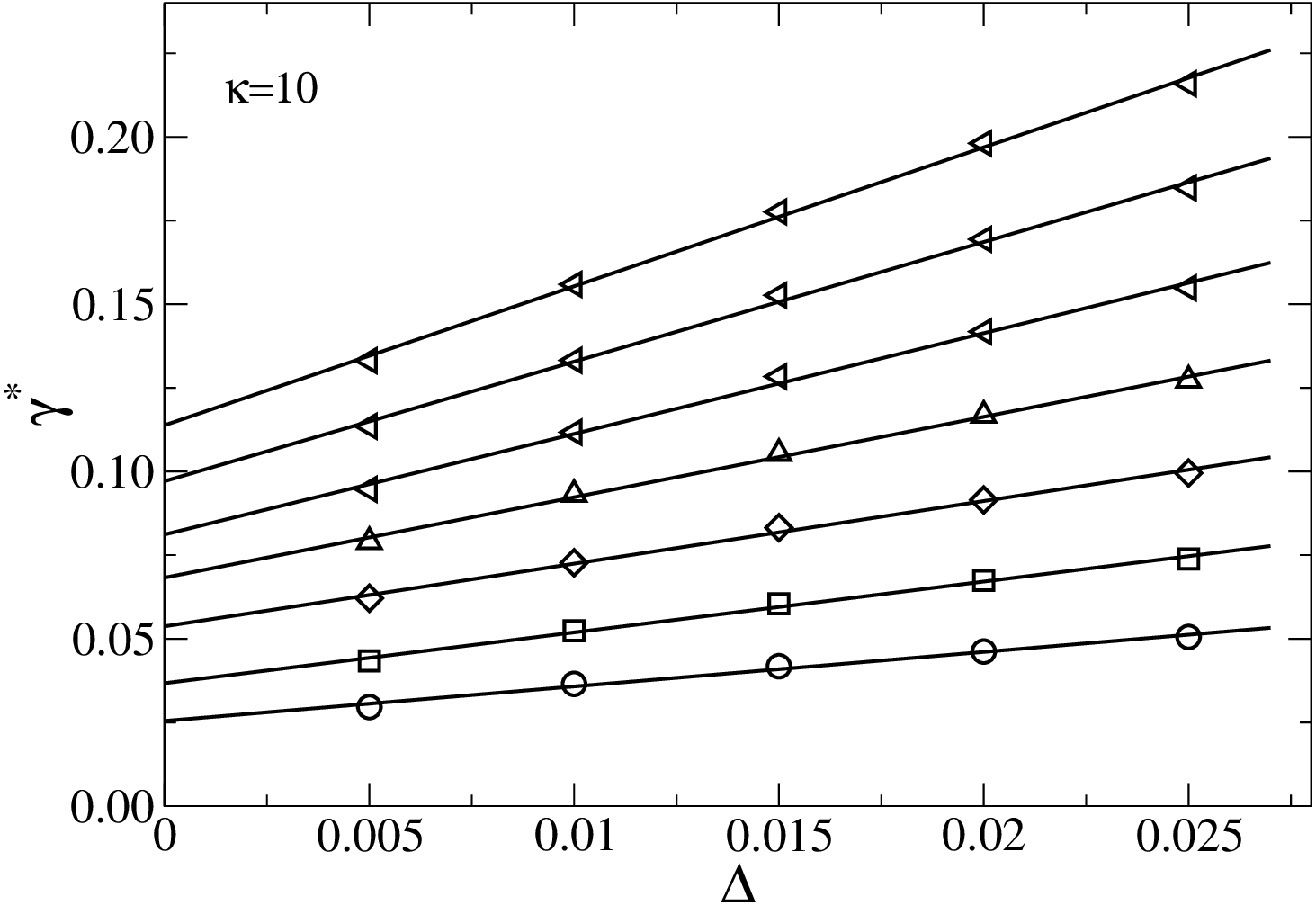}}
\caption{\label{fig3} Surface tension against $\Delta$ for $\kappa=10$ and different temperatures. Upper curves correspond to lower temperatures. Lines correspond to linear regressions.}
\end{figure}

As mentioned in section~\ref{methods}, the accurate evaluation of the surface tension implies determining it as a function of the parameter $\Delta$ and performing an extrapolation for $\Delta \rightarrow 0$. Fig.~\ref{fig3} shows the surface tension as a function of $\Delta$ for $\kappa=10$ and different temperatures. In this figure one can see the linear behavior of the surface tension as a function of $\Delta$ for all studied temperatures. This allows to obtain an accurate value of this property by simply performing a linear extrapolation for $\Delta \rightarrow 0$. It should be noted that the slopes of the regressions never vanish, meaning that a single $\Delta$ value cannot be used for an accurate determination of the surface tension. Moreover, the slopes turn larger for decreasing temperature (see Fig.~\ref{fig3}). Thus, determining the surface tension for the short-range HCAY potential at low temperatures implies obtaining it as a function of $\Delta$ and then performing a careful extrapolation for $\Delta \rightarrow 0$. On the other hand, when comparing the behavior of the surface tension as a function of $\Delta$ for the HCAY fluid to the square well fluid, we notice that the slopes for the HCAY fluid have the opposite sign. This is probably the effect of having a single discontinuity instead of the two corresponding to the square well potential. Anyway, in both cases the linear extrapolation is essential for the accurate evaluation of the surface tension of discontinuous potentials by the virial route. 

\begin{figure}
\resizebox{0.4\textwidth}{!}{\includegraphics{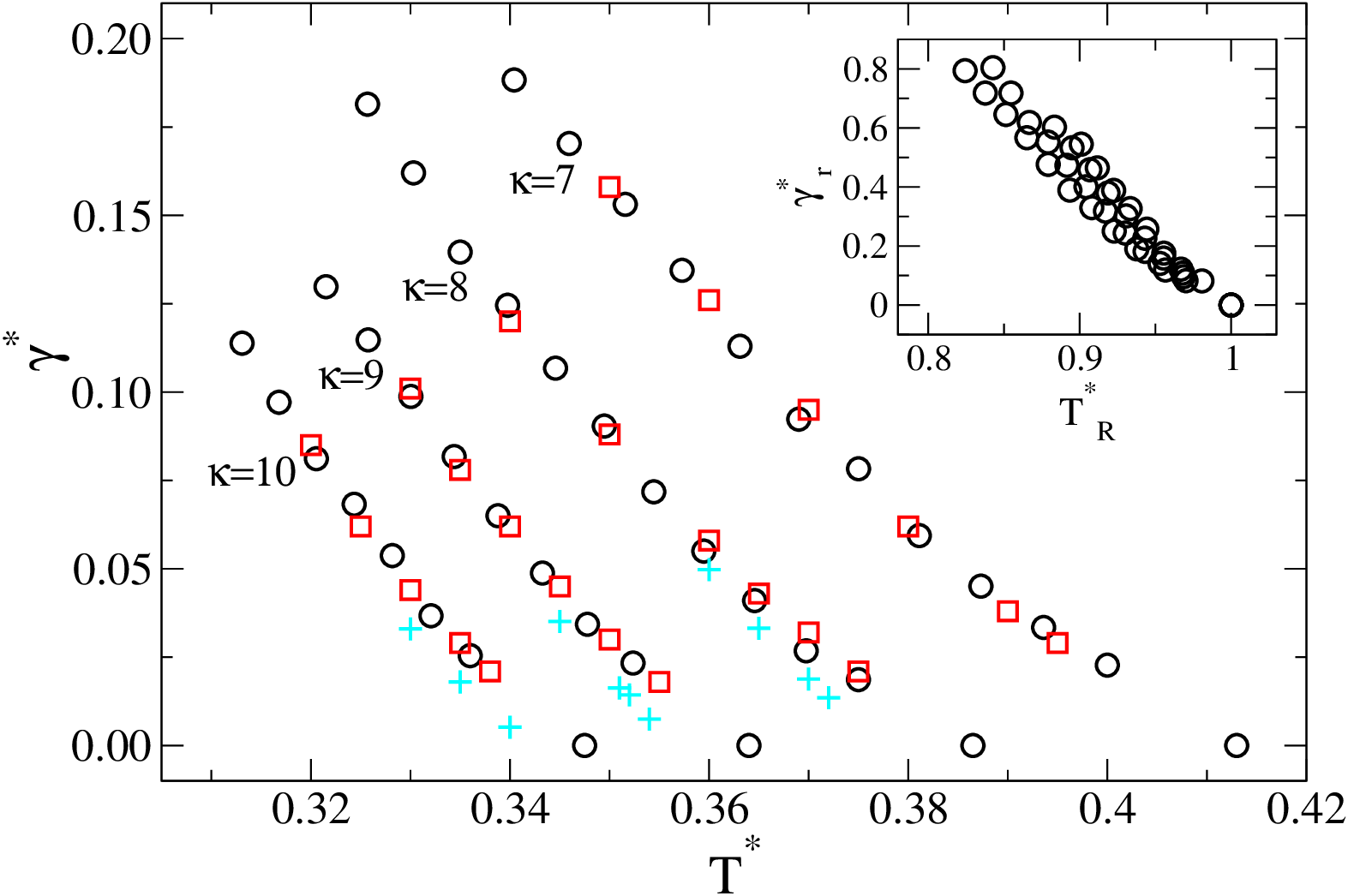}}
\caption{\label{fig4} Surface tension of strong short-range HCAY fluid as a function of
temperature and $\kappa$. Circles are obtained by REMC and squares by the standard MC method. Crosses corresponds to data reported by Singh~\cite{Singh09}. The inset shows the REMC reduced surface tension against the reduced temperature.}
\end{figure}

The results of the surface tension as a function of the temperature for all considered $\kappa$ values are shown in Fig.~\ref{fig4} and given in Table~\ref{Table1}. The curves behavior is very similar to that found with longer interaction ranges, i.~e., the surface tension raises as temperature decreases and the curves shift to the left with increasing $\kappa$. As in previous plots, circles correspond to REMC and squares to the standard MC method. The obtained agreement is remarkable. This provides confidence in the proposed method for the surface tension evaluation~\cite{Alejandre99,Orea03,Orea05} of these particular extremely short-range and discontinuous potentials by the virial method. 

\begingroup
\squeezetable
\begin{table}
\caption{Surface tension values for the strong short-range Yukawa fluids. The subscripts are the estimated errors. } \label{Table1}
\begin{tabular}{||c|c|c|c|c||}
\hline
\hline
& \multicolumn{2}{c|}{Standard MC} & \multicolumn{2}{c||}{REMC} \\
\hline
$\kappa$ & $T^*$ & $\gamma^*$ & $T^*$ & $\gamma^*$ \\
\hline
  & 0.395 & $0.029_{07}$ & 0.4000 & $0.023_{06}$ \\
  & 0.390 & $0.038_{05}$ & 0.3936 & $0.033_{07}$ \\
  & 0.380 & $0.062_{06}$ & 0.3873 & $0.045_{04}$ \\
  & 0.370 & $0.095_{10}$ & 0.3811 & $0.059_{05}$ \\
  & 0.360 & $0.126_{08}$ & 0.3750 & $0.078_{07}$ \\
7 & 0.350 & $0.158_{09}$ & 0.3690 & $0.092_{06}$ \\
  &       &             & 0.3631 & $0.113_{07}$ \\
  &       &             & 0.3573 & $0.134_{06}$ \\
  &       &             & 0.3516 & $0.153_{05}$ \\
  &       &             & 0.3460 & $0.170_{11}$ \\
  &       &             & 0.3404 & $0.188_{08}$ \\
\hline
  & 0.375 & $0.021_{06}$ & 0.3750 & $0.019_{06}$  \\
  & 0.370 & $0.032_{07}$ & 0.3698 & $0.027_{05}$  \\
  & 0.365 & $0.043_{05}$ & 0.3646 & $0.041_{07}$  \\
  & 0.360 & $0.058_{06}$ & 0.3595 & $0.055_{05}$  \\
  & 0.350 & $0.088_{08}$ & 0.3544 & $0.072_{05}$  \\
8 & 0.340 & $0.120_{11}$ & 0.3495 & $0.090_{06}$  \\
  &       &              & 0.3446 & $0.107_{08}$  \\
  &       &              & 0.3398 & $0.125_{06}$  \\
  &       &              & 0.3350 & $0.140_{10}$  \\
  &       &              & 0.3303 & $0.162_{08}$  \\
  &       &              & 0.3257 & $0.182_{09}$  \\
\hline
  & 0.355 & $0.018_{04}$ & 0.3524 & $0.023_{06}$  \\
  & 0.350 & $0.030_{06}$ & 0.3478 & $0.034_{08}$  \\
  & 0.345 & $0.045_{08}$ & 0.3433 & $0.049_{05}$  \\
9 & 0.340 & $0.062_{05}$ & 0.3388 & $0.065_{07}$  \\
  & 0.335 & $0.078_{11}$ & 0.3344 & $0.082_{06}$  \\
  & 0.330 & $0.101_{08}$ & 0.3300 & $0.099_{08}$  \\
  &       &              & 0.3258 & $0.115_{10}$  \\
  &       &              & 0.3215 & $0.130_{09}$  \\
\hline
  & 0.338 & $0.021_{05}$ & 0.3360  & $0.025_{04}$  \\
  & 0.335 & $0.029_{08}$ & 0.3321  & $0.037_{07}$  \\
  & 0.330 & $0.044_{07}$ & 0.3282  & $0.054_{06}$  \\
10& 0.325 & $0.062_{07}$ & 0.3243  & $0.068_{09}$  \\
  & 0.320 & $0.085_{09}$ & 0.3205  & $0.081_{11}$  \\
  &       &              & 0.3168  & $0.097_{05}$  \\
  &       &              & 0.3131  & $0.114_{08}$  \\
\hline
\hline
\end{tabular}
\end{table}
\endgroup

We also included in Fig.~\ref{fig4} the data recently reported by Singh by means of GC-TMMC method~\cite{Singh09}. A comparison to these data reveals a good qualitative agreement. That is, trends show a good match. Nonetheless, our data are systematically above Singh's values, specially close to the critical point where differences are above $30\%$. It should be mentioned that differences are considerably larger than our error bars, which are always smaller than twice the symbol size for both cases (see Table 1). These differences in the surface tension values are despite the fact that a very good match was found between the vapor-liquid phase diagrams obtained by GC-TMMC and the standard MC technique, as it was shown in a previous work~\cite{Pini10}. Our belief is that the GC-TMMC technique would probably yield larger surface tension data for a considerably larger number of MC steps (the author reported a short running time calculation, $48-72 hrs$). Note that we performed $6\times 10^7$ and $4\times 10^7$ steps with the standard MC and the REMC method, respectively, to produce the data shown in Fig.~\ref{fig4}. A $M=12$ REMC run takes approximately a month in a quad core desktop machine. Thus, in both cases, a large number of steps was required. In that sense, the gain of employing REMC instead of a standard MC technique for the surface tension determination is not very remarkable (we expected a much better performance of REMC). A similar finding was obtained by de Miguel~\cite{Miguel08} when comparing the expanded ensemble approach (a clever method also based on the introduction of an extended ensemble and thus related to REMC) with the virial route for obtaining the surface tension.

Previous findings for intermediate and long interaction range of the same pair potential show the surface tension and the coexistence curves yield a master curve when reduced using their critical parameters~\cite{Orea08}. Furthermore, a single master curve was also found for the coexistence data of the strong short-range HCAY~\cite{Pini10}. However, the reduced surface tension, $\gamma^*_r=\gamma/(\rho^{*2/3}_c T^*_c)$, obtained by REMC as a function of the reduced temperature, $T^*_R=T^*/T^*_c$, shown in the inset of Fig.~\ref{fig4} forms an imperfect master curve, where slight deviations are observed. Thus, we cannot assure the obeying of the corresponding states law of the strong short-range HCAY based on this property.

\section{Conclusions}
The presented work fulfills the three purposes stated in the introduction. That is, it shows that the virial route calculation of the surface tension for discontinuous potentials\cite{Alejandre99,Orea03} produces reliable data for strong short-range systems. For this purpose, the surface tension must be obtained as a function of $\Delta$ and then a careful extrapolation for $\Delta \rightarrow 0$ must be implemented. Then, it reveals the utility of the REMC technique to assist the virial route for obtaining data at low temperatures. Finally, it reports new interface tension values for the strong short-range HCAY fluids which contribute to the knowledge on the behavior of these systems for a broader range of thermodynamic conditions. These new data may help testing the results of the existing and forthcoming theoretical approaches.  

\section{Acknowledgments}
The authors thank The Molecular Engineering Program of IMP as well as CONACyT of M\'exico for financial support (projects Nos. D.00406 and Y.00119 SENER-CONACyT).\\


\end{document}